
%
\input phyzzx
\tolerance=1000
\sequentialequations
\def\rl{\rightline}

\def\etal{{\it et. al.}}

\def\t1{{\tilde 1}}

\def\AEF{A.E. Faraggi}

\def\SSM{supersymmetric standard model}
\def\NPB#1#2#3{Nucl. Phys. B{\bf#1} (19#2) #3}
\def\PLB#1#2#3{Phys. Lett. B{\bf#1} (19#2) #3}
\def\PRD#1#2#3{Phys. Rev. D{\bf#1} (19#2) #3}

\REF\FFM{I. Antoniadis \etal, \PLB{231}{89}{65};
 \AEF, D.V. Nanopoulos and K. Yuan, \NPB{335}{90}{347};
 I. Antoniadis, G. K. Leontaris and J. Rizos, \PLB{245}{90}{161};
 \AEF, \PLB{278}{92}{131}; \NPB{387}{92}{239}, hep-th/9208024;
 J.L. Lopez, D.V. Nanopoulos, and K. Yuan, \NPB{399}{93}{654}, hep-th/9203025.}
\REF\PD{\AEF, \NPB{428}{94}{111}.}
\REF\FMH{J.L. Lopez and D.V. Nanopoulos, \NPB{338}{90}{73};
					 \PLB{251}{90}{73};
	 \AEF, \NPB{407}{92}{57}; \AEF~and E. Halyo, \NPB{416}{94}{63}.}
\REF\TOP{\AEF, \PLB{274}{92}{47}.}
\REF\CHSW{P. Candelas, G.T. Horowitz, A. Strominger and E. Witten,
                                                        \NPB{258}{85}{46}.}
\REF\DHVW{L. Dixon, J.A. Harvey, C. Vafa and E. Witten, \NPB{261}{85}{678};
                                                        \NPB{274}{86}{285}.}
\REF\FFF{H. Kawai, D.C. Lewellen, and S.H.-H. Tye, \NPB{288}{87}{1};
         I. Antoniadis, C. Bachas, and C. Kounnas, \NPB{289}{87}{87};
         I. Antoniadis and C. Bachas, \NPB{298}{88}{586}.}
\REF\KS{A. Kagan and S. Samuel, \PLB{284}{92}{289}.}
\REF\FOC{\AEF, \PLB{326}{94}{62}.}
\REF\CCM{B. de Carlos, J.A. Casas and C. Mu\~noz, \NPB{399}{93}{623}.}
\REF\naturalness{A.E. Faraggi and D.V. Nanopoulos, \PRD{48}{93}{3288}.}
\REF\Narain{K.S. Narain, \PLB{169}{86}{41};
            K.S. Narain, M.H. Sarmadi and E. Witten, \NPB{279}{87}{369}.}
\REF\JOL{S. Chaudhuri, G. Hockney, and J. Lykken,
         preprint FERMILAB-PUB-95/349-T, hep-th/9510241.}
\REF\DLNR{H. Dreiner, J.L. Lopez, D.V. Nanopoulos and D. Reiss,
					\NPB{320}{89}{401};
	  				G. Cleaver, hepth/9505080.}
\REF\SSM{L.E. Iba{\~n}ez {\it{et al.}}, \NPB{301}{88}{157};
	 L.E. Iba{\~n}ez {\it{et al.}}, \PLB{191}{87}{282};
	 A. Font {\it{et al.}}, \NPB{331}{90}{421};
         J.A. Casas, E.K. Katehou and C. Mu{\~n}oz, \NPB{317}{89}{171}.}

\singlespace
\rl{UFIFT-HEP-95-29}
\rl{November 1995}
\normalspace
\medskip
\titlestyle{\bf On the origin of three generation
	        free fermionic superstring models{\footnote*{work supported
by DE--FG05--86ER--40272.}}}
\author{Alon E. Faraggi{\footnote\dag{Talk presented at the
			Wakulla Springs workshop on Supersymmetry.}}
			{\footnote\ddag{e--mail address:
					faraggi@phys.ufl.edu}}}
\medskip
\centerline {Department of Physics, University of Florida}
\centerline {Gainesville, FL 32611}
\bigskip
\titlestyle{ABSTRACT}
The three generation superstring models in the free fermionic models
have had remarkable success in describing the real--world. The most
explored models use the NAHE set to obtain three generations and to separate
the hidden and observable sectors. It is of course well known that the full
NAHE set is not required in order to construct three generation free fermionic
models. I argue that all the semi--realistic free fermionic models that have
been constructed
to date correspond to $Z_2\times Z_2$ orbifolds. Thus, the successes of the
semi--realistic free fermionic models, if taken seriously, suggest that
the true string vacuum is a $Z_2\times Z_2$ orbifold with
nontrivial background fields and quantized Wilson lines.

\singlespace
\vskip 0.5cm
\nopagenumbers
\pageno=0
\endpage
\normalspace
\pagenumbers

\centerline{\bf 1. Introduction}

The realistic free fermionic superstring models have had remarkable
success in accounting for the observed low energy physics [\FFM]. Not only do
these models give rise to three chiral generations with the
correct quantum numbers under the Standard Model gauge group, but perhaps more
impressive is their success in providing plausible explanation to various
properties of the observed low energy spectrum, like the stability of the
proton [\PD] and the fermion mass hierarchy [\FMH].
Perhaps the most outstanding success of the realistic free fermionic models
is the correct prediction of the top quark mass [\TOP],
that was obtained in the
context of these models several years prior to the experimental observation
of the top quark by the CDF/D0 collaborations.

The successes of the realistic free fermionic superstring
models may lead one to speculate that this indeed may be the path
that nature has chosen. It should be emphasized that it is
not claimed that one of the string models that were constructed
to date is the string vacuum that nature has chosen. Indeed, such
a claim will require far more elaborate analysis than has been performed
to date. However, the remarkable successes of the free
fermionic superstring models provide evidence that suggest that
the eventually emerging true string vacua will share some of the
basic underlying features of the realistic free fermionic models.

Taking this point of view, it is then of extreme importance to try
to extract what are the basic underlying features of the realistic
free fermionic models. One of the common properties of the realistic
free fermionic superstring models is the fact that they all have three
chiral generations.

It is well known that in the bosonic formulation of compactification
of the heterotic string to four dimensions the number of generations
is related to the Euler characteristic of the compactified manifold
[\CHSW,\DHVW].
The free fermionic formulation [\FFF], however, is formulated at a fixed point
in the compactified space and all the information on the geometry of
the compactified manifold is lost. The bosonic formulation of the
heterotic string
has a great advantage over the fermionic formulation in the
sense that we can continuously deform the parameters of the compactified
space and connect between string vacua that in the fermionic formulation
would appear as distinct models. Thus, an extremely important task is to try
to determine what is the underlying compactification of the realistic
free fermionic models.

In this talk I argue that the underlying compactification of all the
three generation free fermionic superstring models (that have
been constructed to date) is $Z_2\times Z_2$ orbifold compactification.
A very simple realization of this underlying geometry is
achieved with the so called ``NAHE'' set. However, it is well known
that the complete NAHE{\footnote*{This set was first
constructed by Nanopoulos, Antoniadis, Hagelin and Ellis  (NAHE)
in the construction
of  the flipped $SU(5)$.  {\it nahe}=pretty, in
Hebrew.}} set is not required for obtaining
three generations free fermionic models [\KS].
I argue that also in the case
of non-NAHE models, there is an underlying compactification of a
$Z_2\times Z_2$ orbifold.
I propose that the successes of the realistic free fermionic
models, if taken seriously, indicate that
the true string vacuum is a $Z_2\times Z_2$ orbifold with
nontrivial background fields and quantized Wilson lines.

\bigskip
\centerline{\bf 2. Three generation models with the NAHE set}

In the free fermionic formulation of the heterotic string
in four dimensions all the world--sheet
degrees of freedom  required to cancel
the conformal anomaly are represented in terms of free fermions.
For the left--movers one has the
usual space--time fields $X^\mu$, $\psi^\mu$, ($\mu=0,1,2,3$),
and in addition the following eighteen real free fermion fields:
$\chi^I,y^I,\omega^I$  $(I=1,\cdots,6)$, transforming as the adjoint
representation of $SU(2)^6$.
A model in this construction
is defined by a set of boundary condition basis vectors, which
are constrained by the string
consistency requirements.
The basis vectors generate a finite additive group
$\Xi$. The physical states in the Hilbert space, of a given sector
$\alpha\epsilon\Xi$, are obtained by acting on the vacuum
with bosonic, and fermionic operators. For a periodic complex fermion $f$,
there are two degenerate vacua ${\vert +\rangle},{\vert -\rangle}$ ,
annihilated by the zero modes $f_0$ and
${{f_0}^*}$ and with fermion numbers  $F(f)=0,-1$, respectively.
The physical spectrum is obtained by applying
the generalized GSO projections.

The basis vectors of the NAHE set are
$\{{\bf 1},S,b_1,b_2,b_3\}$
with a choice of generalized GSO projections [\naturalness].
The sector ${S}$ generates  $N=4$ space--time supersymmetry, which is broken
to $N=2$ and $N=1$ space--time supersymmetry by $b_1$ and $b_2$, respectively.
The gauge group after the NAHE set is $SO(10)\times
E_8\times SO(6)^3$.
At the level of the NAHE set, each
sector $b_1$, $b_2$ and  $b_3$  give rise to 16 spinorial 16 of $SO(10)$.
The Neveu-Schwarz sector produces massless states that transform as
$(5\oplus\bar5)$ of $SO(10)$ and as singlets of $SO(10)\times E_8$.

The NAHE set divides the internal world--sheet fermions into several groups.
The internal $44$ right--moving  fermionic states
are divided in the following way:
${\bar\psi}^{1,\cdots,5}$ are complex and produce the observable $SO(10)$
symmetry;
${\bar\phi}^{1,\cdots,8}$ are complex and produce the hidden $E_8$ gauge
group;
$\{{\bar\eta}^1,{\bar y}^{3,\cdots,6}\}$, $\{{\bar\eta}^2,{\bar y}^{1,2}
,{\bar\omega}^{5,6}\}$, $\{{\bar\eta}^3,{\bar\omega}^{1,\cdots,4}\}$
 give rise to the three horizontal $SO(6)$ symmetries.
The left--moving $\{y,\omega\}$ states
are divided to,
$\{{y}^{3,\cdots,6}\}$, $\{{y}^{1,2}
,{\omega}^{5,6}\}$, $\{{\omega}^{1,\cdots,4}\}$.
The left--moving $\chi^{12},\chi^{34},\chi^{56}$ states
carry the supersymmetry charges.

An important consequence of the NAHE set is
observed by extending the $SO(10)$
symmetry to $E_6$. Adding
to the NAHE set a vector $X$
with periodic boundary conditions for the set
$\{{{\bar\psi}^{1,\cdots,5}},
{{\bar\eta}^{1,2,3}}\}$, extends the gauge symmetry to
$E_6\times U(1)^2\times SO(4)^3$.
The sectors $(b_j;b_j+X)$, $(j=1,2,3)$
each give eight $27$ of $E_6$. The $(NS;NS+X)$ sector gives in
addition to the vector bosons and spin two states, three copies of
scalar representations in $27+{\bar {27}}$ of $E_6$.

In this model the fermionic states which count the
multiplets of $E_6$ are the internal fermions $\{y,w\vert{\bar y},
{\bar\omega}\}$. The vacuum of the sectors
$b_j$  contains twelve periodic fermions. Each periodic fermion
gives rise to a two dimensional degenerate vacuum $\vert{+}\rangle$ and
$\vert{-}\rangle$ with fermion numbers $0$ and $-1$, respectively.
After applying the
GSO projections, we can write the vacuum of the sector
$b_1$ in combinatorial form
$$\eqalignno{\left[\left(\matrix{4\cr
                                    0\cr}\right)+
\left(\matrix{4\cr
                                    2\cr}\right)+
\left(\matrix{4\cr
                                    4\cr}\right)\right]
\left\{\left(\matrix{2\cr
                                    0\cr}\right)\right.
&\left[\left(\matrix{5\cr
                                    0\cr}\right)+
\left(\matrix{5\cr
                                    2\cr}\right)+
\left(\matrix{5\cr
                                    4\cr}\right)\right]
\left(\matrix{1\cr
                                    0\cr}\right)\cr
+\left(\matrix{2\cr
                                    2\cr}\right)
&\left[\left(\matrix{5\cr
                                    1\cr}\right)+
\left(\matrix{5\cr
                                    3\cr}\right)+
\left(\matrix{5\cr
                                    5\cr}\right)\right]\left.
\left(\matrix{1\cr
                                    1\cr}\right)\right\}&(1)\cr}$$
where
$4=\{y^3y^4,y^5y^6,{\bar y}^3{\bar y}^4,
{\bar y}^5{\bar y}^6\}$, $2=\{\psi^\mu,\chi^{12}\}$,
$5=\{{\bar\psi}^{1,\cdots,5}\}$ and $1=\{{\bar\eta}^1\}$.
The combinatorial factor counts the number of $\vert{-}\rangle$ in
a given state. The two terms in the curly brackets correspond to the two
components of a Weyl spinor.  The $10+1$ in the $27$ of $E_6$ are
obtained from the sector $b_j+X$.
The states which count the multiplicities of $E_6$ are the internal
fermionic states $\{y^{3,\cdots,6}\vert{\bar y}^{3,\cdots,6}\}$.
A similar result is
obtained for the sectors $b_2$ and $b_3$ with $\{y^{1,2},\omega^{5,6}
\vert{\bar y}^{1,2},{\bar\omega}^{5,6}\}$
and $\{\omega^{1,\cdots,4}\vert{\bar\omega}^{1,\cdots,4}\}$
respectively, which suggests that
these twelve states correspond to a six dimensional
compactified orbifold with Euler characteristic equal to 48.

The same model is generated in the
orbifold language
by moding out an $SO(12)$ lattice by a $Z_2\times{Z_2}$
discrete symmetry with standard embedding [\FOC]. The $SO(12)$ lattice
is obtained for special values of the metric and antisymmetric tensor
and at a fixed point in compactification space. The metric is the Cartan
matrix of $SO(12)$ and the antisymmetric tensor is given by $b_{ij}=g_{ij}$
for $i>j$. The sectors $b_1$, $b_2$ and $b_3$ correspond to
the three twisted sectors in the orbifold models and the Neveu--Schwarz sector
corresponds to the untwisted sector.

The reduction to three generations is illustrated in table 1.
In the realistic free fermionic models the vector $X$
is replaced by the vector $2\gamma$ in which $\{{\bar\psi}^{1,\cdots,5},
{\bar\eta}^1,{\bar\eta}^2,{\bar\eta}^3,{\bar\phi}^{1,\cdots,4}\}$
are periodic. This reflects the fact that these models
have (2,0) rather than (2,2) world-sheet supersymmetry.
At the level of the NAHE set we have 48 generations.
One half of the generations is projected because of the vector $2\gamma$.
Each of the three vectors in table 1 acts nontrivially on the degenerate
vacuum of the fermionic states
$\{y,\omega\vert{\bar y},{\bar\omega}\}$ that are periodic in the
sectors $b_1$, $b_2$ and $b_3$ and reduces the combinatorial
factor of Eq. (1) by a half. Thus, we obtain one generation from each sector
$b_1$, $b_2$ and $b_3$.

\bigskip
\centerline{\bf 2. Three generation models without the NAHE set}

In the previous section we saw how three generation free fermionic
models are obtained if the basis contains the full NAHE set,
$\{{\bf 1},S,b_1,b_2,b_3\}$. In these models the connection with the
$Z_2\times Z_2$ orbifold is readily established. The basis vector
$b_3$ is replaced with the basis vector $\xi_1=1+b_1+b_2+b_3$.
The set $\{{\bf 1},S,\xi_1\}$ generates a toroidal compactified model with
$N=4$ supersymmetry and $SO(28)\times E_8$ gauge group. The
boundary condition vectors $b_1$ and $b_2$ the correspond to the
$Z_2\times Z_2$ orbifold twist. The three sectors $b_1$,
$b_2$ and $b_3={\bf 1}+b_1+b_2+\xi_1$
then correspond to three twisted sectors of the
$Z_2\times Z_2$ orbifold model. The reduction to three generations
is achieved by reducing the number of fixed points from each twisted
sector to one, by adding three additional boundary condition basis
vectors. Each subsequent boundary condition basis vector reduce the
number of fixed points from each sector $b_1$, $b_2$ and $b_3$
by a factor of two. At the same time the gauge group is broken
to one of the maximal subgroups of $SO(10)$. Many other desirable
phenomenological properties, like doublet--triplet splitting,
can be achieved for appropriate assignment of the boundary conditions
in the additional boundary condition basis vectors.

The free fermionic formulation is formulated at a point
in the moduli space that may be prefered from a dynamical point of view.
The reason being that the free fermionic formulation is formulated near the
self--dual point in the compactification space. Studies of the effective
moduli potential within the context of dynamical supersymmetry
breaking by gaugino condensates, indeed suggest that moduli VEVS of the
order of the self--dual radius minimize the effective moduli potential [\CCM].
The structure exhibited by the NAHE set is then seen to be very robust
in getting three generation models with very desirable phenomenological
properties.

These properties of the realistic free fermionic models suggest that
it may be that the true string vacuum is in the close vicinity of these
models. An extremely important question is then to ask what are the properties
of these models that are truly model independent. While it may be
naive to expect
that one specific string model will turn out to be the true string vacuum,
it is reasonable
to expect that perhaps we will be able to guess the correct string
compactification, or the correct neighborhood of the true string vacuum.

In the language of the free fermionic models the low energy
phenomenological properties are related to the choices of boundary
condition basis vectors and generalized GSO phases.
We can then ask which choices of basis vectors and GSO phases, in this
limited class of models, are
necessary to obtain certain desirable phenomenological properties.
One such phenomenological criteria is the requirement of three chiral
generations.

While the NAHE set provides an elegant way for obtaining three
chiral generations,
it is of course well known that the full NAHE set is not required
for constructing three generation free fermionic models.
The first published example of such a model was published in Ref. [\KS].
This model is shown in table 3 (with a slight change of notation).
In this model the sectors that may produce chiral generations are the
sectors $b_1$, $b_2$ and $b_3$. The degenerate vacuum of the fermionic
zero modes can be represented similar to Eq. (1). The internal fermionic
states can also be divided into the same groups as with the NAHE set.
Each sector $b_j$ then gives rise to 16 generations. This number
is then reduced by the choices of boundary conditions in the
remaining boundary condition basis vectors. It is easy to see from table 2
that in this model the sector $b_1$ and $b_2$ give rise to one and two
generations respectively.
{}From the table we observe that the degeneracy of the vacuum due to the
real fermions in the sector $b_1$ is removed completely, while in the
sector $b_2$ a double degeneracy remains. The sector $b_3$ does not obey
the chirality condition of Ref. [\naturalness]
and therefore gives rise only to
non-chiral matter. The chirality condition of Ref. [\naturalness] states that
to obtain from a given sector, $b_j$, chiral $16$ representation of $SO(10)$
we need a second vector, $b_k$, with
$\{{\psi^\mu},{{\bar\psi}_{1{\cdots}5}}\}$ periodic in both vectors and
the intersection between the remaining boundary conditions is empty.
If this condition is not satisfied then, as  long as the $SO(10)$ symmetry
is not broken, a given vector $b_j$ will give an equal number of
$16$ and  ${\bar {16}}$ and
thus will not contribute to the net number of generations.
The vector $b_k$ is not necessarily a basis vectors but must exist as a
combination of basis vectors. Therefore in this model the sector $b_3$
produces one $16$ and one ${\overline{16}}$ representation of $SO(10)$.

This model is an example how three generation can be obtained without
the full NAHE set. Of course, there may exist many other possibilities.
For example, as shown in Ref. [\KS] it is possible to add the vector
$\alpha$ with
$\{y^5,\omega^5,y^6,\omega^6,\vert{\bar y}^5,{\bar\omega}^5,
   {\bar y}^6,{\bar\omega}^6,{\bar\eta}^1,{\bar\phi}^{6,\cdots,8}\}$
periodic and the remaining boundary conditions antiperiodic to the basis of
table 3.
With this basis vector, the combination $b_1+\alpha$ projects the
$\overline{16}$ from the sector $b_3$ and reduces the double degeneracy of
the generations from the sector $b_2$. Thus, with basis vector $\alpha$
this model contains one generation from each sector $b_1$, $b_2$ and $b_3$.

The important question, however, is whether all these three generation
models possess some common structure.

The answer, of course, is that all these models are related to
$Z_2\times Z_2$ orbifolds. The orbifold moding however
does not act on a torodially compactified model with $N=4$
space--time supersymmetry and
$SO(12)\times E_8\times E_8$ or $SO(12)\times SO(16)\times SO(16)$
gauge group, as in the case of the models that contain the full NAHE set.
In the non--NAHE models, the sectors $\xi_1$ and $\xi_2$ that produce
the spinorial of $SO(16)$ in the observable and hidden $E_8$s are not
present, therefore the space--time gauge group of the torodially
compactified model is $SO(44)$. This gauge symmetry is obtained
in the bosonic formulation for appropriate choices of the metric,
antisymmetric tensor and Wilson lines [\Narain].

In the Kagan--Samuel model above
the basis vectors $\{{\bf 1},S\}$ generates a torodially
compactified model with $N=4$ and $SO(44)$ gauge group. The
basis vectors $b_1$ and $b_2$ correspond to the $Z_2\times Z_2$ orbifold
twisting. The third twisted sector of the orbifold model is still
present, and is the combination ${\bf 1}+b_1+b_2$.
In this model the vector $\xi_2$ which
generates the spinorial representation of $SO(16)$ in the adjoint
representation of the hidden $E_8$ gauge group is absent. Consequently,
the states from the third twisted sector, ${\bf 1}+b_1+b_2$,
are not massless states.

All the three generation
free fermionic models that have been constructed to date contain the
basis vectors $\{{\bf 1},S\}$. The basis vector $S$ produces $N=4$ space--time
supersymmetry. The only way to reduce the number of supersymmetry
generators that are produced by this basis vector, is by moding by a
$Z_2\times Z_2$ twist. Therefore, the connection between the three
generation free fermionic models and $Z_2\times Z_2$ orbifold compactification
is in fact expected. By free fermions we confine ourselves to conformal
field theory blocks that are either complex fermions or Ising model operators.
Recently, it was found that three generation models can also be obtained if
one relaxes this constraint [\JOL]. In the models of Ref. [\JOL], the vector
$S$ is used to generate the space--time supersymmetry. Consequently, in this
models, the $Z_2\times Z_2$ structure is preserved in the supersymmetric
sector, while it is relaxed in the bosonic sector. These models
make use of more complicated conformal solutions that do not correspond
to free fermions.

\bigskip
\centerline{\bf 4. Conclusion}

In this talk, I discussed the correspondence between three generation
free fermionic models and orbifold compactification. All the three generation
free fermionic models that have been constructed to date correspond
to $Z_2\times Z_2$ orbifold compactifications. Other compactifications
may be constructed in the free fermionic formulation by using different
supersymmetry generators [\DLNR] (It is of course possible that
these SUSY generators can also lead to three generation models).
If the successes of the realistic free fermionic models are to be taken
seriously, then they indicate that the true string vacuum is related
to a $Z_2\times Z_2$ orbifold compactification with nontrivial background
fields. Out of the plethora of orbifold compactifications [\SSM],
$Z_2\times Z_2$ orbifolds, therefore, deserve special attention.

\refout

\vfill
\eject

\input tables.tex
\nopagenumbers
\magnification=1000
\tolerance=1200

{\hfill
{\begintable
\  \ \|\ ${\psi^\mu}$ \ \|\ $\{{\chi^{12};\chi^{34};\chi^{56}}\}$\ \|\
{${\bar\psi}^1$, ${\bar\psi}^2$, ${\bar\psi}^3$,
${\bar\psi}^4$, ${\bar\psi}^5$, ${\bar\eta}^1$,
${\bar\eta}^2$, ${\bar\eta}^3$} \ \|\
{${\bar\phi}^1$, ${\bar\phi}^2$, ${\bar\phi}^3$, ${\bar\phi}^4$,
${\bar\phi}^5$, ${\bar\phi}^6$, ${\bar\phi}^7$, ${\bar\phi}^8$} \crthick
$\alpha$
\|\ 0 \|
$\{0,~0,~0\}$ \|
1, ~~1, ~~1, ~~0, ~~0, ~~0 ,~~0, ~~0 \|
1, ~~1, ~~1, ~~1, ~~0, ~~0, ~~0, ~~0 \nr
$\beta$
\|\ 0 \| $\{0,~0,~0\}$ \|
1, ~~1, ~~1, ~~0, ~~0, ~~0, ~~0, ~~0 \|
1, ~~1, ~~1, ~~1, ~~0, ~~0, ~~0, ~~0 \nr
$\gamma$
\|\ 0 \|
$\{0,~0,~0\}$ \|
{}~~$1\over2$, ~~$1\over2$, ~~$1\over2$, ~~$1\over2$,
{}~~$1\over2$, ~~$1\over2$, ~~$1\over2$, ~~$1\over2$ \| $1\over2$, ~~0, ~~1,
{}~~1,
{}~~$1\over2$,
{}~~$1\over2$, ~~$1\over2$, ~~0 \endtable}
\hfill}
\smallskip
{\hfill
{\begintable
\  \ \|\
${y^3y^6}$,  ${y^4{\bar y}^4}$, ${y^5{\bar y}^5}$,
${{\bar y}^3{\bar y}^6}$
\ \|\ ${y^1\omega^6}$,  ${y^2{\bar y}^2}$,
${\omega^5{\bar\omega}^5}$,
${{\bar y}^1{\bar\omega}^6}$
\ \|\ ${\omega^1{\omega}^3}$,  ${\omega^2{\bar\omega}^2}$,
${\omega^4{\bar\omega}^4}$,  ${{\bar\omega}^1{\bar\omega}^3}$  \crthick
$\alpha$ \|
1, ~~~0, ~~~~0, ~~~~0 \|
0, ~~~0, ~~~~1, ~~~~1 \|
0, ~~~0, ~~~~1, ~~~~1 \nr
$\beta$ \|
0, ~~~0, ~~~~1, ~~~~1 \|
1, ~~~0, ~~~~0, ~~~~0 \|
0, ~~~1, ~~~~0, ~~~~1 \nr
$\gamma$ \|
0, ~~~1, ~~~~0, ~~~~1 \|\
0, ~~~1, ~~~~0, ~~~~1 \|
1, ~~~0, ~~~~0, ~~~~0  \endtable}
\hfill}
\smallskip
\parindent=0pt
\hangindent=39pt\hangafter=1
\baselineskip=18pt

{{\it Table 1.} A three generations model with the NAHE set.
The choice of generalized GSO coefficients is:
${c\left(\matrix{b_j\cr
                                    \alpha,\beta,\gamma\cr}\right)=
-c\left(\matrix{\alpha\cr
                                    1\cr}\right)=
c\left(\matrix{\alpha\cr
                                    \beta\cr}\right)=
-c\left(\matrix{\beta\cr
                                    1\cr}\right)=
c\left(\matrix{\gamma\cr
                                    1,\alpha\cr}\right)=
-c\left(\matrix{\gamma\cr
                                    \beta\cr}\right)=
-1}$ (j=1,2,3), with the others specified by modular invariance and space--time
supersymmetry. }
\vskip 2.5cm

\vfill
\eject

{\hfill
{\begintable
\  \ \|\ ${\psi^\mu}$ \ \|\ $\{{\chi^{12};\chi^{34};\chi^{56}}\}$  \ \|\
{${\bar\psi}^1$, ${\bar\psi}^2$, ${\bar\psi}^3$,
${\bar\psi}^4$, ${\bar\psi}^5$, ${\bar\eta}^1$,
${\bar\eta}^2$, ${\bar\eta}^3$} \ \|\
{${\bar\phi}^1$, ${\bar\phi}^2$, ${\bar\phi}^3$, ${\bar\phi}^4$,
${\bar\phi}^5$, ${\bar\phi}^6$, ${\bar\phi}^7$, ${\bar\phi}^8$} \crthick
${\bf 1}$
\|\ 1 \| $\{1,~1,~1\}$  \|
1, ~~1, ~~1, ~~1, ~~1, ~~1, ~~1, ~~1 \|
1, ~~1, ~~1, ~~1, ~~1, ~~1, ~~1, ~~1 \nr
${S}$
\|\ 1 \|
$\{1,~1,~1\}$  \|
0, ~~0, ~~0, ~~0, ~~0, ~~0, ~~0, ~~0 \|
0, ~~0, ~~0, ~~0, ~~0, ~~0, ~~0, ~~0 \crthick
${b_1}$
\|\ 1 \| $\{1,~0,~0\}$  \|
1, ~~1, ~~1, ~~1, ~~1, ~~1, ~~0, ~~0 \|
0, ~~0, ~~0, ~~0, ~~0, ~~0, ~~0, ~~0 \nr
${b_2}$
\|\ 1 \|
$\{0,~1,~0\}$  \|
1, ~~1, ~~1, ~~1, ~~1, ~~0, ~~1, ~~0 \|
0, ~~0, ~~0, ~~0, ~~0, ~~0, ~~0, ~~0 \nr
${b_3}$
\|\ 1 \|
$\{0,~1,~0\}$  \|
1, ~~1, ~~1, ~~1, ~~1, ~~0, ~~1, ~~0 \|
0, ~~0, ~~0, ~~0, ~~0, ~~0, ~~0, ~~0  \crthick
${P}$
\|\ 0 \| $\{0,~0,~0\}$  \|
0, ~~0, ~~0, ~~0, ~~0, ~~0, ~~0, ~~0 \|
0, ~~0, ~~0, ~~0, ~~1, ~~1, ~~1, ~~1 \nr
${\alpha}$
\|\ 0 \|
$\{0,~0,~0\}$  \|
{}~~$1\over2$, ~~$1\over2$, ~~$1\over2$, ~~$1\over2$,
{}~~$1\over2$, ~~$1\over2$, ~~$1\over2$, ~~$1\over2$ \| $1\over2$,
{}~~$1\over2$, ~~$1\over2$, ~~$1\over2$, ~~0, ~~0, ~~1, ~~1
\endtable}
\hfill}
\smallskip
{\hfill
{\begintable
\  \ \|\
${y^3y^4}$,  ${y^5{\bar y}^5}$, ${y^6{\bar y}^6}$,
${{\bar y}^3{\bar y}^4}$
\ \|\ ${y^1{\bar y}^6}$,  ${y^2{\bar y}^2}$,
${\omega^5{\bar\omega}^5}$,
${{\bar\omega}^6{\bar\omega}^6}$
\ \|\ ${\omega^1{\bar\omega}^1}$,  ${\omega^2{\bar\omega}^2}$,
${\omega^3{\bar\omega}^3}$,  ${{\omega}^4{\bar\omega}^4}$ \crthick
${\bf1}$ \|
1, ~~~1, ~~~~1, ~~~~1 \|
1, ~~~1, ~~~~1, ~~~~1 \|
1, ~~~1, ~~~~1, ~~~~1 \nr
${S}$ \|
0, ~~~0, ~~~~0, ~~~~0 \|
0, ~~~0, ~~~~0, ~~~~0 \|
0, ~~~0, ~~~~0, ~~~~0 \crthick
${b_1}$ \|
1, ~~~1, ~~~~1, ~~~~1 \|
0, ~~~0, ~~~~0, ~~~~0 \|
0, ~~~0, ~~~~0, ~~~~0 \nr
${b_2}$ \|
0, ~~~0, ~~~~0, ~~~~0 \|\
1, ~~~1, ~~~~1, ~~~~1 \|
0, ~~~0, ~~~~0, ~~~~0 \nr
${b_3}$ \|
0, ~~~1, ~~~~1, ~~~~0 \|\
0, ~~~0, ~~~~0, ~~~~0 \|
1, ~~~1, ~~~~0, ~~~~0 \crthick
${P}$ \|
0, ~~~1, ~~~~0, ~~~~0 \|
0, ~~~0, ~~~~1, ~~~~0 \|
0, ~~~0, ~~~~0, ~~~~0 \nr
${\alpha}$ \|
0, ~~~0, ~~~~0, ~~~~1 \|\
1, ~~~0, ~~~~0, ~~~~0 \|
1, ~~~0, ~~~~0, ~~~~0 \endtable}
\hfill}
\smallskip
\parindent=0pt
\hangindent=39pt\hangafter=1
\baselineskip=18pt

{{\it Table 2.} The three generations
${SU(5)\times U(1)}$ Kagan--Samuel model.
The choice of GSO phases is:
${c\left(\matrix{$S$\cr
                                    {\bf 1}, b_j\cr}\right)=
 -c\left(\matrix{$S$\cr
                                    P,\alpha\cr}\right)=
  c\left(\matrix{b_1\cr
                                    b_2,b_3\cr}\right)=
  c\left(\matrix{b_2\cr
                                    b_3\cr}\right)=
 -c\left(\matrix{b_j\cr
                                    P\cr}\right)=
 -c\left(\matrix{b_1\cr
                                    \alpha\cr}\right)=
1}$ (j=1,2,3) and ${c\left(\matrix{b_2,b_3\cr
                                    \alpha\cr}\right)=
i}$, with the others specified by modular invariance and space--time
supersymmetry (The notation and GSO phases may be different from KS).
This model does not contain the full NAHE set. }

\vskip 2cm

\bye

\end